\documentclass[twocolumn,preprintnumbers,pra,hyperref]{revtex4}
\usepackage{amssymb}
\usepackage{amsfonts}
\usepackage{amsmath}
\usepackage{graphicx}

\begin{document}

\title{Damping law of photocount distribution in a dissipative channel%
\thanks{{\small This work was supported by the National Natural Science
Foundation of China (Grant Nos. 11175113 and 11264018), and the Young
Talents Foundation of Jiangxi Normal University.}}}
\author{Hong-Yi Fan$^{1}$, Sen-Yue Lou$^{1}$ and Li-Yun Hu$^{2\dag }$}
\affiliation{$^{1}${\small Department of Physics,\ Ningbo University, Ningbo 315211, P.
R. China}\\
$^{2}${\small Department of physics, Jiangxi Normal University, Nanchang
330022, P. R. China }\\
$\dag ${\small Email: hlyun@jxnu.edu.cn.}}

\begin{abstract}
{\small For a dissipative channel governed by the master equation of density
operator }$d\rho /dt=\kappa \left( 2a\rho a^{\dagger }-a^{\dagger }a\rho
-\rho a^{\dagger }a\right) ,${\small \ we find that photocount distribution
formula at time }$t,${\small \ }$p\left( n,t\right) =Tr\left\{ \rho \left(
t\right) \mathbf{\colon }\left( \xi a^{\dagger }a\right) ^{n}e^{-\xi
a^{\dagger }a}/n!\colon \right\} ,${\small \ becomes }$p\left( n,t\right) =Tr%
\left[ \rho \left( 0\right) \mathbf{\colon }\left( \xi e^{-2\kappa
t}a^{\dagger }a\right) ^{n}e^{-\xi e^{-2\kappa t}a^{\dagger }a}/n!\colon %
\right] ,${\small \ as if the quantum efficiency }$\xi ${\small \ of the
detector becomes }$\xi e^{-2\kappa t}${\small .\ This law greatly simplifies
the theoretical study of photocount distribution for quantum optical field.}
\end{abstract}

\maketitle

In nature, syetems we concerned usually are surrounded by thermo reservoir.
Decoherence is an important topic in the fields of quantum optics, quantum
computing and quantum information processing. Decoherence of a quantum
optical field can be judged by photocounting its final state comparing with
its initial state. In this paper we discuss{\small \ }time-dissipation of
photocount distribution in a dissipative reservoir. We hope to know photon
counting law for $\rho (t)$, when the initial density operator $\rho
_{0}=\rho (t=0)$ evolves with time in a photon-loss channel. The
corresponding master equation is \cite{r1}%
\begin{equation}
\frac{d\rho }{dt}=\kappa \left( 2a\rho a^{\dagger }-a^{\dagger }a\rho -\rho
a^{\dagger }a\right) ,  \label{r1}
\end{equation}%
where $\kappa $ is the rate of dissipation, $\left[ a,a^{\dagger }\right] =1$%
.\ We shall deduce a new damping formula of photon counting after an optical
quantum field passing through a photon-loss channel.

\emph{Converting }$\rho $\emph{'s master equations into pure state's
evolution equations. }In order to solve equation Eq.(\ref{r1}), instead of
converting the operator equation into its $c$-number equation by the Wigner
function approach or by \textit{P}-representation approach, we introduce
thermo entangled state representation \cite{r2}
\begin{equation}
\left \vert \tau \right \rangle =\exp \left( -\frac{1}{2}|\tau |^{2}+\tau
a^{\dagger }-\tau ^{\ast }\tilde{a}^{\dagger }+a^{\dagger }\tilde{a}%
^{\dagger }\right) \left \vert 0\tilde{0}\right \rangle ,  \label{r2}
\end{equation}%
which is complete, to make up the pure state $\left \vert \rho
\right
\rangle $,
\begin{equation}
\left \vert \rho \right \rangle \equiv \rho \left \vert I\right \rangle ,%
\text{ }\left \vert I\right \rangle \equiv \left \vert \tau =0\right \rangle
=e^{a^{\dagger }\tilde{a}^{\dagger }}\left \vert 0\tilde{0}\right \rangle
=\sum_{n=0}^{\infty }\left \vert n\tilde{n}\right \rangle ,  \label{r3}
\end{equation}%
where $\tilde{a}^{\dagger }$ is a fictitious mode accompanying the real
photon creation operator $a^{\dagger },$ $\left \vert \tilde{0}%
\right
\rangle $ is annihilated by $\tilde{a},$ $\left[ \tilde{a},\tilde{a}%
^{\dagger }\right] =1,$ $\left \vert n\tilde{n}\right \rangle =\left(
a^{\dagger }\tilde{a}^{\dagger }\right) ^{n}\left \vert 0\tilde{0}%
\right
\rangle /n!,\ $ $\left[ a,\tilde{a}^{\dagger }\right] =0$. From Eq.(%
\ref{r3}) it is easy to see that%
\begin{eqnarray}
a\left \vert I\right \rangle &=&\tilde{a}^{\dagger }\left \vert I\right
\rangle ,  \notag \\
a^{\dagger }\left \vert I\right \rangle &=&\tilde{a}\left \vert I\right
\rangle ,  \notag \\
\left( a^{\dagger }a\right) ^{n}\left \vert I\right \rangle &=&\left( \tilde{%
a}^{\dagger }\tilde{a}\right) ^{n}\left \vert I\right \rangle ,  \label{r4}
\end{eqnarray}%
which indicates that the transformation relation between the real mode and
the fictitious mode. Acting the both sides of Eq.(\ref{r1}) on the state $%
\left \vert I\right \rangle ,$ we have%
\begin{equation}
\frac{d}{dt}\left \vert \rho \right \rangle =\kappa \left( 2a\rho a^{\dagger
}-a^{\dagger }a\rho -\rho a^{\dagger }a\right) \left \vert I\right \rangle .
\label{r5}
\end{equation}%
Using Eq.(\ref{r4}) and noting that $\rho $ is defined in the real space
which is commutative with the fictitious operators, we can convert Eq.(\ref%
{r5}) into
\begin{equation}
\frac{d}{dt}\left \vert \rho \right \rangle =\kappa \left( 2a\tilde{a}%
-a^{\dagger }a-\tilde{a}^{\dagger }\tilde{a}\right) \left \vert \rho \right
\rangle .  \label{r6}
\end{equation}%
Its formal solution is given by%
\begin{equation}
\left \vert \rho \right \rangle =\exp \{ \kappa t\left( 2a\tilde{a}%
-a^{\dagger }a-\tilde{a}^{\dagger }\tilde{a}\right) \} \left \vert \rho
_{0}\right \rangle ,  \label{r7}
\end{equation}%
where $\left \vert \rho _{0}\right \rangle \equiv \rho _{0}\left \vert
I\right \rangle $. Thus by\ introducing a fictitious mode $\tilde{a}%
^{\dagger }$ to be a counterpart of the system mode, one can convert density
operators' master equations into evolution equations of pure states, which
leads to the formal solution (\ref{r7}).

On the other hand, noticing that the commutative relation $\left[
(a^{\dagger }a+\tilde{a}^{\dagger }\tilde{a})/2,a\tilde{a}\right] =-a\tilde{a%
},$ thus we can use the operator identity%
\begin{equation}
e^{\lambda \left( A+\sigma B\right) }=e^{\lambda A}\exp \left[ \sigma \left(
1-e^{-\lambda \tau }\right) B/\tau \right] ,  \label{r8}
\end{equation}%
which is valid for $\left[ A,B\right] =\tau B$, to decompose the exponential
operator in Eq.(\ref{r7}) as \cite{r3}
\begin{equation}
e^{-2\kappa t(\frac{a^{\dagger }a+\tilde{a}^{\dagger }\tilde{a}}{2}-a\tilde{a%
})}=e^{-\kappa t\left( a^{\dagger }a+\tilde{a}^{\dagger }\tilde{a}\right)
}e^{Ta\tilde{a}},  \label{r9}
\end{equation}%
where $T=1-e^{-2\kappa t}.$ On substituting Eq.(\ref{r9}) into Eq.(\ref{r7})
yields
\begin{eqnarray}
\left\vert \rho \left( t\right) \right\rangle  &=&\exp \left[ -\kappa
t\left( a^{\dagger }a+\tilde{a}^{\dagger }\tilde{a}\right) \right] \exp %
\left[ Ta\tilde{a}\right] \left\vert \rho _{0}\right\rangle   \notag \\
&=&\exp \left[ -\kappa t\left( a^{\dagger }a+\tilde{a}^{\dagger }\tilde{a}%
\right) \right] \sum_{n=0}^{\infty }\frac{T^{n}}{n!}a^{n}\tilde{a}^{n}\rho
_{0}\left\vert I\right\rangle .  \label{r10}
\end{eqnarray}%
Noticing that the system mode is commutative with the fictitious operators,
and the transition relation shown in Eq.(\ref{r4}), we see that

\begin{equation}
\tilde{a}^{n}\rho _{0}\left \vert I\right \rangle =\rho _{0}\tilde{a}%
^{n}\left \vert I\right \rangle =\rho _{0}a^{\dagger n}\left \vert I\right
\rangle ,  \label{r11}
\end{equation}%
and%
\begin{eqnarray}
&&\exp \left[ -\kappa t\tilde{a}^{\dagger }\tilde{a}\right] a^{n}\rho
_{0}a^{\dagger n}\left \vert I\right \rangle  \notag \\
&=&a^{n}\rho _{0}a^{\dagger n}\exp \left[ -\kappa t\tilde{a}^{\dagger }%
\tilde{a}\right] \left \vert I\right \rangle  \notag \\
&=&a^{n}\rho _{0}a^{\dagger n}\exp \left[ -\kappa ta^{\dagger }a\right]
\left \vert I\right \rangle ,  \label{r12}
\end{eqnarray}%
thus we have%
\begin{eqnarray}
\rho \left( t\right) \left \vert I\right \rangle &=&e^{-\kappa ta^{\dagger
}a}\sum_{n=0}^{\infty }\frac{T^{n}}{n!}a^{n}\rho _{0}a^{\dagger n}e^{-\kappa
t\tilde{a}^{\dagger }\tilde{a}}\left \vert I\right \rangle  \notag \\
&=&\sum_{n=0}^{\infty }\frac{T^{n}}{n!}e^{-\kappa ta^{\dagger }a}a^{n}\rho
_{0}a^{\dagger n}e^{-\kappa ta^{\dagger }a}\left \vert I\right \rangle .
\label{r13}
\end{eqnarray}%
Comparing with the standard form of the operator-sum representation \cite{r4}%
,

\begin{equation}
\rho \left( t\right) =\sum_{m=0}^{\infty }M_{m}\rho _{0}M_{m}^{\dagger }
\label{r14}
\end{equation}%
we see that the Kraus operator for this damping process is given by
\begin{equation}
M_{m}=\sqrt{\frac{T^{m}}{m!}}e^{-\kappa ta^{\dagger }a}a^{m},  \label{r15}
\end{equation}%
and we can prove the unitrity $\sum_{m=0}^{\infty }M_{m}^{\dagger }M_{m}=1$.

\emph{Damping formula of Photocount distribution in a dissipative channel.}
Now we discuss how does the photocount distribution formula of optical
fields evolve in the photon-loss channel?

The quantum mechanical photon counting formula was first derived by Kelley
and Kleiner \cite{r5}. As shown in Ref. \cite{r5,r6,r7,r8,r9} the
probability distribution $\mathfrak{p}\left( n\right) $ of registering $n$
photoelectrons in time interval $\Delta \tau $ is given by

\begin{equation}
\mathfrak{p}\left( n\right) =\mathtt{Tr}\left\{ \rho \mathbf{\colon }\frac{%
\left( \xi a^{\dagger }a\right) ^{n}}{n!}e^{-\xi a^{\dagger }a}\colon
\right\} ,  \label{r16}
\end{equation}%
where $\mathbf{\colon }\colon $ denotes normal ordering, $\xi \propto \Delta
\tau $ is called the quantum efficiency (a measure) of the detector, $%
\mathbf{\rho }\left( 0\right) $ is the single-mode density operator of the
light field concerned at initial time. According to the operator-sum
representation at time $t$ in Eq.(\ref{r14}), we have%
\begin{eqnarray}
\mathfrak{p}\left( n,t\right) &=&\mathtt{Tr}\left\{ \rho \left( t\right)
\mathbf{\colon }\frac{\left( \xi a^{\dagger }a\right) ^{n}}{n!}e^{-\xi
a^{\dagger }a}\colon \right\}  \notag \\
&=&\mathtt{Tr}\left\{ \rho _{0}\hat{O}\left( t\right) \right\} ,  \label{r17}
\end{eqnarray}%
where we have set%
\begin{equation}
\hat{O}\left( t\right) =\sum_{m=0}^{\infty }M_{m}^{\dagger }\mathbf{\colon }%
\frac{\left( \xi a^{\dagger }a\right) ^{n}}{n!}e^{-\xi a^{\dagger }a}\colon
M_{m}.  \label{r18}
\end{equation}%
Substituting Eq.(\ref{r15}) into Eq.(\ref{r18}) and using the following
operator identities: $\mathbf{\colon }e^{-\xi a^{\dagger }a}\colon =\exp %
\left[ a^{\dagger }a\ln \left( 1-\xi \right) \right] =\left( 1-\xi \right)
^{a^{\dagger }a}$ and $e^{\lambda a^{\dagger }a}a^{\dagger }e^{-\lambda
a^{\dagger }a}=a^{\dagger }e^{\lambda }$, $e^{\lambda a^{\dagger
}a}ae^{-\lambda a^{\dagger }a}=ae^{-\lambda }$, we can put Eq.(\ref{r18})
into the following form%
\begin{eqnarray}
&&\hat{O}\left( t\right)  \notag \\
&=&\sum_{m=0}^{\infty }\frac{T^{m}}{m!}a^{\dagger m}e^{-\kappa ta^{\dagger
}a}\mathbf{\colon }\frac{\left( \xi a^{\dagger }a\right) ^{n}}{n!}e^{-\xi
a^{\dagger }a}\mathbf{\colon }e^{-\kappa ta^{\dagger }a}a^{m}  \notag \\
&=&\sum_{m=0}^{\infty }\frac{T^{m}}{m!}a^{\dagger m}e^{-\kappa ta^{\dagger
}a}\frac{\xi ^{n}a^{\dagger n}}{n!}\left( 1-\xi \right) ^{a^{\dagger
}a}a^{n}e^{-\kappa ta^{\dagger }a}a^{m}  \notag \\
&=&\frac{\xi ^{n}}{n!}e^{-2n\kappa t}\sum_{m=0}^{\infty }\frac{T^{m}}{m!}%
a^{\dagger n+m}e^{-\kappa ta^{\dagger }a}e^{a^{\dagger }a\ln \left( 1-\xi
\right) }e^{-\kappa ta^{\dagger }a}a^{n+m}  \notag \\
&=&\frac{\xi ^{n}}{n!}e^{-2\kappa tn}\sum_{m=0}^{\infty }\frac{T^{m}}{m!}%
a^{\dagger n+m}\mathbf{\colon }e^{\left[ e^{\left[ \ln \left( 1-\xi \right)
-2\kappa t\right] }-1\right] a^{\dagger }a}\mathbf{\colon }a^{n+m}.
\label{r19}
\end{eqnarray}%
Further using the property that $a$ and $a^{\dagger }$ are commute within
the normal ordering symbol $:$ $:$, we then remember $T=1-e^{-2\kappa t}$ to
have%
\begin{eqnarray}
&&\hat{O}\left( t\right)  \notag \\
&=&\frac{\xi ^{n}}{n!}e^{-2\kappa tn}a^{\dagger n}\colon \sum_{m=0}^{\infty }%
\frac{\left( Ta^{\dagger }a\right) ^{m}}{m!}e^{\left[ e^{\left[ \ln \left(
1-\xi \right) -2\kappa t\right] }-1\right] a^{\dagger }a}\colon a^{n}  \notag
\\
&=&\frac{\left( \xi e^{-2\kappa t}\right) ^{n}}{n!}a^{\dagger n}\colon \exp
\{-\xi e^{-2\kappa t}a^{\dagger }a\}\colon a^{n}  \notag \\
&=&\colon \frac{\left( \xi e^{-2\kappa t}a^{\dagger }a\right) ^{n}}{n!}%
e^{-\xi e^{-2\kappa t}a^{\dagger }a}\colon .  \label{r20}
\end{eqnarray}%
Substituting Eq.(\ref{r20}) into (\ref{r17}) yields%
\begin{equation}
\mathfrak{p}\left( n,t\right) =\mathtt{Tr}\left\{ \rho _{0}\mathbf{\colon }%
\frac{\left( \xi e^{-2\kappa t}a^{\dagger }a\right) ^{n}}{n!}e^{-\xi
e^{-2\kappa t}a^{\dagger }a}\colon \right\} .  \label{r21}
\end{equation}%
Comparing Eq.(\ref{r21}) with Eq.(\ref{r16}) we obtain the law of damping
formula of photocount distribution in a dissipative channel, that is%
\begin{eqnarray}
\mathfrak{p}\left( n,0\right) &=&\mathtt{Tr}\left\{ \rho \left( 0\right)
\mathbf{\colon }\frac{\left( \xi a^{\dagger }a\right) ^{n}}{n!}e^{-\xi
a^{\dagger }a}\colon \right\}  \notag \\
\left. \rightarrow \right. \mathfrak{p}\left( n,t\right) &=&\mathtt{Tr}%
\left\{ \rho _{0}\mathbf{\colon }\frac{\left( \xi e^{-2\kappa t}a^{\dagger
}a\right) ^{n}}{n!}e^{-\xi e^{-2\kappa t}a^{\dagger }a}\colon \right\} ,
\label{r22}
\end{eqnarray}%
namely,\ this formula exhibits the law as if the quantum efficiency $\xi $
of the detector becomes $\xi e^{-2\kappa t}$. So from Eq.(\ref{r21}) one can
see that the photocount distribution is related to the decay of quantum
efficiency $\xi ^{\prime }=\xi e^{-2\kappa t}$. This is our main result.

\emph{Example.} Now we check the formula through an example. When the
initial state is a pure number state, $\rho \left( 0\right) =\left\vert
m\right\rangle \left\langle m\right\vert ,$ using the following operator
identities $a^{\dagger m}a^{m}=N\left( N-1\right) \cdots \left( N-m+1\right)
$, where $N=a^{\dagger }a$, then the photocount distribution is%
\begin{eqnarray}
&&\mathfrak{p}\left( n,0\right)  \notag \\
&=&\left\langle m\right\vert \mathbf{\colon }\frac{\left( \xi a^{\dagger
}a\right) ^{n}}{n!}e^{-\xi a^{\dagger }a}\mathbf{\colon }\left\vert
m\right\rangle  \notag \\
&=&\frac{\xi ^{n}}{n!}\sum_{l=0}^{\infty }\frac{\left( -\xi \right) ^{l}}{l!}%
\left\langle m\right\vert a^{\dagger n+l}a^{n+l}\left\vert m\right\rangle
\notag \\
&=&\frac{\xi ^{n}}{n!}\sum_{l=0}^{\infty }\frac{\left( -\xi \right) ^{l}}{l!}%
\left\langle m\right\vert N\left( N-1\right) \cdots \left( N-n-l+1\right)
\left\vert m\right\rangle  \notag \\
&=&\sum_{l=0}^{m-n}\frac{\left( -\xi \right) ^{l}}{l!}\frac{\xi ^{n}}{n!}%
\frac{m!}{\left( m-n-l\right) !}  \notag \\
&=&\frac{m!}{n!\left( m-n\right) !}\xi ^{n}\left( 1-\xi \right) ^{m-n},
\label{r23}
\end{eqnarray}%
which indicates that the probability of a photon being counted during the
period $\Delta \tau $ is the quantum efficiency $\xi $, thus the probability
of counting $n$ out of $m$ photons is proportional to the probability of
counting $n$ photons $\xi ^{n}$ times the probability of counting $m-n$
photons $\left( 1-\xi \right) ^{m-n}$. Then according to the new formula Eq.(%
\ref{r22}) we immediately have the probability distribution of optical filed
$\left\vert m\right\rangle $ after damping at time $t,$ i.e.,
\begin{eqnarray}
\mathfrak{p}\left( n,t\right) &=&\mathtt{Tr}\left\{ \rho _{0}\mathbf{\colon }%
\frac{\left( \xi e^{-2\kappa t}a^{\dagger }a\right) ^{n}}{n!}e^{-\xi
e^{-2\kappa t}a^{\dagger }a}\mathbf{\colon }\right\}  \notag \\
&=&\frac{m!\left( \xi e^{-2\kappa t}\right) ^{n}}{n!\left( m-n\right) !}%
\left( 1-\xi e^{-2\kappa t}\right) ^{m-n}.  \label{r24}
\end{eqnarray}%
Now we check it through another approach. We see that using (\ref{r14})-(\ref%
{r15}) the initial state $\left\vert m\right\rangle \left\langle
m\right\vert $ evolves into a binomial state,%
\begin{equation}
\left\vert m\right\rangle \left\langle m\right\vert \rightarrow e^{-2m\kappa
t}\sum_{l=0}^{m}\frac{m!\left( e^{2\kappa t}-1\right) ^{l}}{l!\left(
m-l\right) !}\left\vert m-l\right\rangle \left\langle m-l\right\vert ,
\label{r25}
\end{equation}%
then using Eq.(\ref{r16}) and Eq.(\ref{r23}) we have
\begin{eqnarray}
&&\mathfrak{p}\left( n,t\right)  \notag \\
&=&\sum_{l=0}^{m}\frac{m!\left( e^{2\kappa t}-1\right) ^{l}}{l!\left(
m-l\right) !e^{2m\kappa t}}\left\langle m-l\right\vert \mathbf{\colon }\frac{%
\left( \xi a^{\dagger }a\right) ^{n}}{n!}e^{-\xi a^{\dagger }a}\colon
\left\vert m-l\right\rangle  \notag \\
&=&e^{-2m\kappa t}\sum_{l=0}^{m-n}\frac{m!\left( e^{2\kappa t}-1\right) ^{l}%
}{l!n!\left( m-n-l\right) !}\xi ^{n}\left( 1-\xi \right) ^{m-l-n}  \notag \\
&=&\frac{m!\xi ^{n}e^{-2m\kappa t}}{n!\left( m-n\right) !}\sum_{l=0}^{m}%
\binom{m-n}{m-n-l}\left( e^{2\kappa t}-1\right) ^{l}\left( 1-\xi \right)
^{m-n-l}  \notag \\
&=&\frac{m!\left( \xi e^{-2\kappa t}\right) ^{n}}{n!\left( m-n\right) !}%
\left( 1-\xi e^{-2\kappa t}\right) ^{m-n},  \label{r26}
\end{eqnarray}%
which is the same as Eq.(\ref{r24}). Thus our new formula is correct.

In summary, we conclude that for a dissipative system characteristic of
amplitude damping, the photocount distribution at any time can be considered
as the one at initial time with the time evolution quantum efficiency $\xi
e^{-2\kappa t},$ shown in Eq.(\ref{r22}). This law greatly simplifies the
theoretical study of photocount distribution for quantum optical field. In
addition, in Ref.\cite{r10}, we have derived two new quantum-mechanical
photocount formulas which build the relation from Wigner function and
Q-function to photocount distribution. Thus our result can be applied
directly to the case.

\textbf{Acknowledgements:} This work was supported by the National Natural
Science Foundation of China (Grant Nos. 11175113 and 11264018), and the
Young Talents Foundation of Jiangxi Normal University.

\end{document}